# Demonstration of a Tilted-Pulse-Front Pumped Plane-Parallel Slab Terahertz Source

Priyo Syamsul Nugraha,[1,2] Gergő Krizsán,[1] Csaba Lombosi,[1] László Pálfalvi,[3] György Tóth,[3] Gábor Almási,[3] József András Fülöp,[1,2,4*] and János Hebling[1,2,3]

[1]*Szentágothai Research Centre, University of Pécs, 7624 Pécs, Hungary*
[2]*MTA-PTE High-Field Terahertz Research Group, 7624 Pécs, Hungary*
[3]*Institute of Physics, University of Pécs, 7624 Pécs, Hungary*
[4]*ELI-ALPS, ELI-Hu Nonprofit Ltd., 6720 Szeged, Hungary*
*Corresponding author: fulop@fizika.ttk.pte.hu

**A new type of tilted-pulse-front pumped terahertz (THz) source has been demonstrated, which is based on a LiNbO$_3$ plane-parallel slab with an echelon structure on its input surface. Single-cycle pulses of 1 µJ energy and 0.30 THz central frequency have been generated with 5×10$^{-4}$ efficiency from such a source. One order-of-magnitude increase in efficiency is expected by pumping a cryogenically cooled echelon of increased size and thickness with a Ti:sapphire laser. The use of a plane-parallel nonlinear optical crystal slab enables straightforward scaling to high THz pulse energies and to produce a symmetric THz beam with uniform pulse shape for good focusability and high field strength.**

The study and control of materials with extremely strong THz fields [1, 2], or the acceleration and manipulation of electrons [3] and protons [4] are emerging applications which require THz sources with unprecedented parameters. Besides the pulse energy, an excellent focusability is also essential to achieve the highest possible field strengths.

Optical rectification of ultrashort laser pulses with tilted pulse front in lithium niobate (LN) [5] has become a standard technique for efficient THz generation. Conventionally, a prism-shaped LN crystal is used with a large wedge angle equal to the pulse-front tilt (63°). Such a source geometry results in a nonuniform pump propagation length across the beam, which can lead to a spatially varying interaction length for THz generation [6]. This negatively affects the THz beam quality and, consequently, the focusability, thereby limiting the achievable field strength. Lateral beam (and eventually waveform) nonuniformity is especially problematic in high-energy THz sources [7], where a large-diameter pump beam is needed.

Different approaches have been proposed to mitigate limitations of tilted-pulse-front pumped THz sources. The contact-grating technique has been proposed to eliminate imaging errors present in conventional setups utilizing a grating and an imaging lens or telescope [8]. Whereas a plane-parallel contact-grating THz source with uniform interaction length could be successfully demonstrated in ZnTe semiconductor [9], the realization of a practical source in LN turned out to be very challenging [10, 11]. The technical requirements on the contact grating can be somewhat relaxed by combining a conventional external pulse-front-tilting setup with a contact grating [12]. However, even in such hybrid setups a prism-shaped LN crystal needs to be used with ~30° wedge angle. This is still sufficiently large to cause strong transversal nonuniformity across cm-sized beams.

Recently, we have proposed a modified hybrid approach to provide uniform interaction length across large pump and THz beams [13]. The setup uses a plane-parallel LN slab as the nonlinear medium, which is equipped with an echelon structure on its input surface. Inside the LN slab, a segmented tilted pulse front is formed with an average tilt angle as required by phase matching. We note that a reflective echelon has been used earlier for tilting the average pump pulse front for efficient THz generation in LN, but this approach still requires to use a prism-shaped LN crystal with the same wedge angle as in the conventional setup [14,15].

In this work we demonstrate the hybrid tilted-pulse-front pumped THz source which utilizes a plane-parallel LN slab with an echelon structure. This new technology opens up the way for constructing efficient sources of high-energy THz pulses with excellent focusability for high field strengths.

The investigated hybrid-type setup [13] is a combination of the conventional scheme, containing a diffraction grating and imaging optics, and a nonlinear material with an echelon profile on its entrance surface (nonlinear echelon slab, NLES, Fig. 1). Pump pulses of 200 fs pulse duration and 1030 nm central wavelength were delivered by a cryogenically cooled Yb:CaF$_2$ regenerative amplifier operating at 1 kHz repetition rate. Up to about 2.5 mJ pump pulse energy was used in the experiment. At the pump wavelength used and at room temperature, a pulse-front tilt of about 63° is required for phase matching in LN. In case of the NLES, this pulse-front tilt angle needs to be produced in air, at the entrance of the crystal. This is in contrast to the conventional setup where a larger angle of 77° is needed in air, which is then reduced according to $\tan(\gamma_{LN}) = \tan(\gamma_{air})/n_g$ to 63° by entering into the LN prism [16, 17]. Here, $n_g = 2.215$ is the group index of LN at the 1030 nm pump wavelength. The smaller tilt angle is advantageous for reducing imaging errors of the grating-lens system [6].

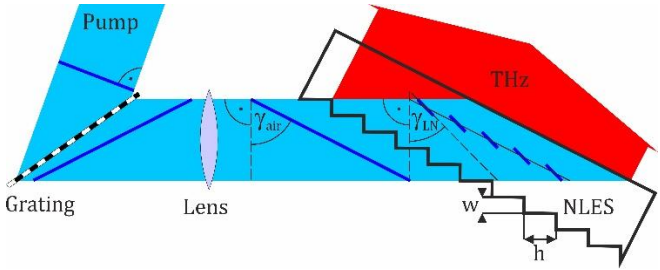

Fig. 1. Scheme of the experimental setup. The pump pulse fronts are indicated at different positions in the setup. Angular dispersion of the pump is not shown. Inside the NLES, the pump pulse front is segmented.

When the pump beam with the pulse-front tilt angle of $\gamma_{air} = 63°$ in air enters the NLES, the local tilt angle is reduced to $\gamma_{LN} = 42°$ (Fig. 1), according to $\tan(\gamma_{LN}) = \tan(\gamma_{air})/n_g$. However, the average pulse-front tilt angle across the pump spot remains unchanged. This leads to the formation of a segmented pulse front [13].

A transmission grating with 1600 mm$^{-1}$ line density (Light Smith, T-1600-1030S) was used at an incidence angle of 55.5°, close to the Littrow angle. To ensure the necessary pulse-front tilt at the crystal, the imaging lens (a near-infrared achromat with 25 cm focal length) had to provide a magnification of about 1.4 times. We note that a more optimal design would involve demagnification, rather than magnification, as then the pump intensity would be higher at the crystal than at the grating. However, we had no appropriate grating to build such a setup. Grating damage was indeed limiting the useful pump intensity in our case. The pump spot size at the NLES, perpendicularly to the pump propagation direction, was 5.5 mm × 5.1 mm (horizontal × vertical, at $1/e^2$ of the peak intensity).

A prototype LN NLES device has been manufactured by diamond milling (Kugler GmbH., Salem, Germany). The dimensions of the slab were 5 mm × 8 mm along and perpendicularly to the steps, respectively. The slab thickness was $L = 3$ mm. The width of the echelon steps, through which the pump beam entered the slab, was $w = 50$ μm (Fig. 1). The height of the steps along the pump propagation direction was $h = 92$ μm. Such a step geometry is consistent with a pulse-front tilt angle of about 62°, required for velocity matching at a cryogenic temperature (100 K), rather than at room temperature. For the sake of simplicity, in this proof-of-principle experiment the NLES was used at room temperature.

The optic axis (Z-axis) of the LN slab was parallel to the (vertical) echelon lines. The slab was Y-cut. The polarization of the pump beam was perpendicular to the plane of Fig. 1. The generated THz beam had the same polarization direction. The THz pulse energy has been measured by a calibrated pyroelectric detector (THZ 20, Sensor- und Lasertechnik). THz waveforms have been measured by electro-optic sampling (EOS) in a [110]-cut 1-mm thick GaP using conventional ellipsometric balanced detection. A small fraction of the 200-fs pump pulses was used for the sampling.

Perspectival-view optical microscope images of the echelon structure are shown in Fig. 2. The surface flatness along the entrance step faces of width $w$ were estimated from interferometric measurements to be about $\lambda/10$ peak-to-peak at the pump wavelength used here, and the rms roughness was 25 nm. It could also be deduced from microscope inspection that the step edges of the echelon profile were rounded (see Fig. 2b). This resulted in an estimated reduction of the effective area by about 15% to 30%.

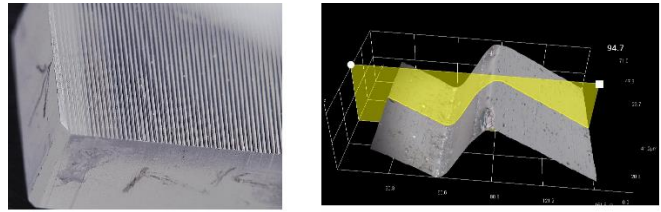

Fig. 2. (a) Perspective view of the prototype echelon slab structure taken at 20× magnification. (b) Reconstructed surface topology model of a single step of the echelon structure (700× magnification). The images were recorded by a Hirox RH-2000 digital microscope by Emilien Leonhardt from Hirox Europe.

Figure 3 shows the measured THz pulse energy (left axis) and the pump-to-THz energy conversion efficiency (right axis) as functions of the pump pulse energy (bottom axis) and the pump intensity (top axis). THz pulses with up to 1.0 μJ energy have been generated with 1.9 mJ of pump reaching the input side of the NLES. The efficiency was 5.1×10$^{-4}$. A nearly quadratic increase of the THz energy and a corresponding linear increase of the efficiency can be observed up to about 1.0 mJ pump energy (25 GW/cm$^2$ intensity). At higher pump energy (intensity) saturation is observed. The useful pump energy (intensity), and consequently the THz energy, has been limited by the damage of the pulse-front-tilting grating, as mentioned above.

For comparison, the obtained THz generation efficiency was by a factor of three smaller for the NLES setup than for a conventional tilted-pulse-front pumped setup consisting of a reflection grating and an achromat lens, used for producing 77° tilt angle of the pump intensity front before the prism-shaped LN crystal with a wedge angle of 63°. According to a simple estimation, however, from this factor of three the small size of the NLES is responsible for a factor of two. If the size of the pump beam can not be much larger than the thickness of the NLES, on a large part of the crystal, the effective generation length will be much shorter than the crystal thickness (see the red area inside the NLES in Fig. 1). This problem can be eliminated by using crystals and pump beams with much larger transversal sizes than the thickness of the NLES. However, in this proof-of-principle experiment, from the 8 mm full width of the NLES crystal only along a 2.4 mm broad part was the THz generation length equal to the 3 mm crystal thickness. The rounded part of the echelon steps at the edges (as shown in Fig. 2b) could also cause some efficiency decrease of the NLES setup compared to an ideal case. However, the effect of this distortion is probably less significant than the effect of the small NLES and pump beam transversal sizes.

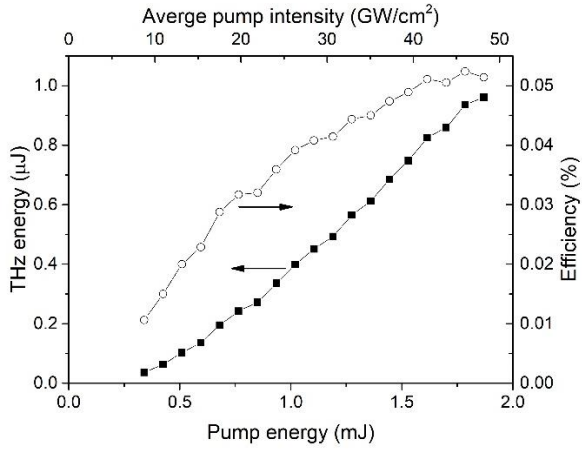

Fig. 3. Measured THz pulse energy and THz generation efficiency as functions of the pump energy and intensity.

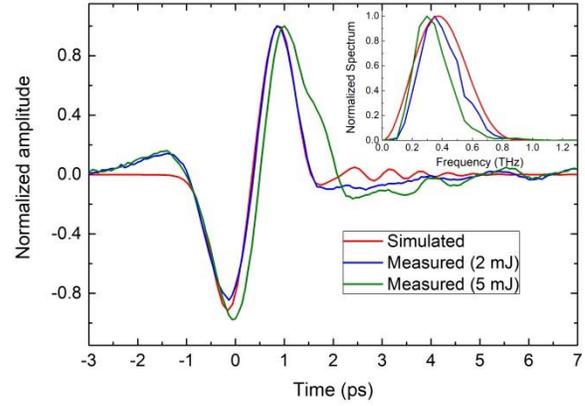

Fig. 4. Measured (at two different pump pulse energies) and simulated THz pulse waveforms and their intensity spectra (inset).

Expectedly, the efficiency achievable experimentally can be improved in several ways. (i) Cryogenic cooling of the NLES reduces the absorption in the THz range and can result in a 2.5× to 3.0× increase of the efficiency [7,18,19]. (ii) As mentioned above, using a significantly larger NLES can result in a 2× increase of the efficiency. (iii) A larger crystal length $L$ (5 mm instead of 3 mm) and a sufficiently large aperture can lead to still further enhancement of the efficiency by a factor of 1.3× [18]. (iv) Because of the smaller diffraction of the segmented pump beam for a shorter wavelength of 800 nm, using a Ti:sapphire laser instead of an Yb-based one can result in a further 40% increase of the efficiency. Taken together, improvements (i) to (iv) can increase the efficiency by about one order of magnitude, reaching $\eta = 0.5\,\%$.

Examples of THz waveforms measured at different pump energies are shown in Fig. 4. These measurements have been carried out with a slightly modified setup, where a 250-cm focal length cylindrical lens have been used for imaging, in combination with a 2-m focal length cylindrical lens. This was used for the reduction of the vertical beam size at the NLES for enabling higher pump intensity without the risk of grating damage. The measured waveforms are nearly single-cycle, in reasonably good accordance with result of numerical simulation, also shown in Fig. 4. The absence of significant ringing on the trailing part can be an important advantage when these pulses are used in particle acceleration setups [20]. The intensity spectra calculated from the measured and simulated waveforms can be seen in the inset of Fig. 4. The simulated spectrum reaches its maximum at 0.38 THz. The spectral peaks of the measured THz pulses are at somewhat lower frequencies. The 0.3 – 0.4 THz mean frequency is suitable for particle acceleration applications.

Increasing the pump energy from 2 mJ to 5 mJ results in the appearance of a clear shoulder on the trailing part of the waveform and a shift of the spectral intensity maximum from 0.35 THz to 0.30 THz. Besides these changes, the saturation of the efficiency can be observed. These observations are clear indications of nonlinear effects. Previously, nonlinear effects such as self-phase-modulation, cascading effect, combination of group-velocity-dispersion and angular dispersion, and stimulated Raman scattering were simulated in conventional tilted-pulse-front pumped setups [21,22]. Another possible nonlinear effect is shift current generation by the THz pulse, which was investigated in LN crystal, but not in the LN THz source [23]. Since the numerical model used in this work does not take into account these nonlinear effects, it can not explain the observed efficiency saturation and waveform and spectral distortions. However, we would like to emphasize that the investigation of the nonlinear effects is less complicated for a plane-parallel crystal than for a prism-shaped LN crystal of 63° wedge angle, as it was the case in the previous investigations.

In conclusion, a new type of tilted-pulse-front pumped THz source has been demonstrated, which is based on a $LiNbO_3$ plane-parallel slab with an echelon structure on its input surface. Single-cycle pulses of 1.0 µJ energy and 0.30 THz central frequency have been generated with $5.1 \times 10^{-4}$ efficiency from such a source. One order-of-magnitude increase in efficiency is expected by pumping a cryogenically cooled echelon of increased size and thickness with a Ti:sapphire laser. The use of a plane-parallel nonlinear optical crystal slab enables straightforward scaling to high THz pulse energies. The pump pulse duration and the THz absorption and dispersion are uniform across the THz beam profile. This advantage, together with the reduced pump beam imaging errors [13], enables to produce a symmetric THz beam with uniform pulse shape for excellent focusability and high field strength. Such a source can be especially suitable to generate low THz frequencies for THz-driven particle acceleration and other applications.

Finally, it is useful to notice that, because of the reduced necessary tilt angle and corresponding angular dispersion of the pre-tilt generation setup, a single transmission grating (instead of a grating-lens combination, see Fig. 1) could be used in front of the nonlinear echelon slab for not larger than 15 mm pump beam diameter and not shorter than 200 fs pump pulse duration.

**Funding.** National Research, Development and Innovation Office (NKFIH) (125808); European Social Fund (EFOP-3.6.2-16-2017-00005: Ultrafast physical processes in atoms, molecules, nanostructures and biology structures).

**Acknowledgment**. We thank Aladár Czitrovszky and Attila Tibor Nagy from the Wigner Research Centre for Physics (Budapest, Hungary) for the help in surface characterization of the NLES sample using a Zygo 7100 optical surface profiler. We also thank the Grimas Kft. for granting access to the Hirox RH-2000 digital microscope.